\documentclass[preprintnumbers,nofootinbib,superscriptaddress,11pt]{revtex4}
\usepackage[dvipdfmx]{graphicx} 
\usepackage{amsmath,amssymb,amsfonts,bm} 

    \def\d{\delta}     \def\th{\theta}     \def\m{\mu} \def\n{\nu}        \def\t{\tau}       

\def\dg{\dagger}  \def\nn{\nonumber}

\newcommand{\meV}{ {\rm meV} }

\newcommand{\lsp}{ \left ( } \newcommand{\rsp}{ \right ) }   \newcommand{\To}{\Rightarrow}   \newcommand{\Getsto}{\Leftrightarrow}

\newcommand{\vev}[1]{ \langle {#1} \rangle }
\def\abs#1{\left| #1\right|}

\newcommand{\Row}[3]{ \begin{pmatrix} #1 & #2 & #3 \end{pmatrix} }
\newcommand{\Column}[3]{ \begin{pmatrix} #1 \\ #2 \\ #3 \end{pmatrix} }

\newcommand{\Diag}[3]{ \begin{pmatrix} #1 & 0 & 0 \\ 0 & #2 & 0 \\ 0 & 0 & #3 \\\end{pmatrix}}

\begin{document}

\title{\Large Chiral perturbative analysis for an almost massless neutrino  \\
in the type-I seesaw mechanism }

\preprint{STUPP-22-260}
\author{Masaki J. S. Yang}
\email{mjsyang@mail.saitama-u.ac.jp}
\affiliation{Department of Physics, Saitama University, 
Shimo-okubo, Sakura-ku, Saitama, 338-8570, Japan}


\begin{abstract} 

In this paper, we perform chiral perturbative analysis of an approximate lepton number symmetry associated with a sufficiently light neutrino in the type-I seesaw mechanism. 
For the Dirac mass matrix $m_{D} = (\bm A \, , \bm B \, , \bm C)$, linearly independent components of $\bm C$ from $\bm A$ and $\bm B$ are treated as  symmetry-breaking parameters. 
A deviation in the eigenvector of the massless mode $\d \bm u$ 
occurs in the first-order perturbation and 
the lightest mass $m_{1 \, \rm or \, 3} \propto \det m_{D}^{2} / M_{3}$ emerges in the second-order. 
By solving the perturbation theory,  
we obtained specific expressions of $\d \bm u$ and $m_{1 \, \rm or \, 3}$. 
As a result, two complex parameters in $m_{D}$ are bounded to some extent from the eigenvector. These constraints are associated with the chiral symmetry and are not susceptible to renormalization. 

\end{abstract} 

\maketitle

\section{Introduction}

The type-I seesaw mechanism \cite{Minkowski:1977sc,GellMann:1980v,Yanagida:1979as, Mohapatra:1979ia} is one of the key subjects in the study of lepton mixing and mass matrix of neutrinos $m_{\n}$. 
While flavor symmetries of $m_{\n}$ have been actively discussed \cite{Ishimori:2010au, Xing:2015fdg},  there are many studies on textures that explore the structure of mass matrices themselves. 
In particular, in the (constrained) sequential dominance \cite{King:1998jw, King:1999cm, King:1999mb, King:2002nf, Antusch:2004gf, Antusch:2004re,  Antusch:2006cw, Antusch:2010tf, King:2005bj, Antusch:2007dj, Antusch:2011ic, Antusch:2013wn, Bjorkeroth:2014vha}, the smallness of the lightest neutrino mass $m_{1 \, \rm or \, 3}$ is  explained by the mass of the heaviest right-handed neutrino $M_{3}$. 

A $U(1)_{L}$ lepton number symmetry is restored in the limit of zero mass for the lightest neutrino \cite{Wyler:1982dd, Petcov:1984nz,  Branco:1988ex}. 
Relations between such a massless neutrino and flavor symmetries have also been studied to some extent \cite{Joshipura:2013pga,Joshipura:2014pqa, King:2016pgv}. 
In particular, Ref.~\cite{Adhikari:2010yt} discussed  general constraints of this symmetry.

In this paper, we explore slightly broken $U(1)_{L}$ symmetry by a very light neutrino 
in the type-I seesaw mechanism.
%
%
Chiral perturbative analysis for finite $m_{1 \, \rm or \, 3}$ shows that 
two complex parameters of the Dirac mass matrix are bounded to some extent by the eigenvectors of the lightest neutrino mass. 

This paper is organized as follows. 
The next section gives a discussion of the $U(1)_{L}$ lepton number symmetry. 
In Sec.~3, we discuss a chiral perturbative analysis of $m_{\n}$ in several methods. 
The final section is devoted to a summary. 

\section{The massless lightest neutrino and $U(1)_{L}$ lepton number symmetry}%

In this section, we consider general constraints  by the $U(1)_{L}$ symmetry associated with only one massless neutrino in the type-I seesaw mechanism. 
A similar analysis exists in Ref.~\cite{Adhikari:2010yt}.
The Dirac mass matrix $m_{D}$ and the symmetric Majorana mass matrix $M_{R}$ of the right-handed neutrinos $\n_{Ri}$ are defined as
\begin{align}
m_{D} =
\begin{pmatrix}
A_1 & B_1 & C_1 \\
A_2 & B_2 & C_2 \\
A_3 & B_3 & C_3 \\
\end{pmatrix} 
\equiv (\bm A \, , \bm B \, , \bm C) \, , 
~~~
M_{R} =
\begin{pmatrix}
 M_{11} & M_{12} & M_{13} \\
 M_{12} & M_{22} & M_{23} \\
 M_{13} & M_{23} & M_{33} \\
\end{pmatrix} .
\label{MR}
\end{align}
These matrix elements $A_{i}, B_{i}, C_{i}$ and $M_{ij}$ are general complex parameters. 
In the type-I seesaw mechanism,  
the mass matrix of light neutrinos $m_{\n}$ is given by
\begin{align}
m_{\n} =  m_{D} M_{R}^{-1} m_{D}^{T} \, . 
\end{align}

To discuss the situation where the lightest neutrino mass $m_{1 \, \rm or \, 3}$ is very light, we first consider the massless limit $m_{1 \, \rm or \, 3} = 0$.
The singular value decomposition (SVD) of $m_{\n}$ is defined as 
\begin{align}
m_{\n} = U m_{\n}^{\rm diag} U^{T} \equiv  (\bm u \, , \bm v \, , \bm w) \Diag{m_{1}}{m_{2}}{m_{3}} (\bm u \, , \bm v \, , \bm w)^{T} \, , 
\end{align}
where $\{ \bm u \, , \bm v \, , \bm w \}$  is a three-dimensional orthonormal basis that constitutes the unitary matrix $U$. 

The massless lightest neutrino $m_{1 \, \rm or \, 3} = 0$ leads to the normal hierarchy (NH) or  the inverted hierarchy (IH), respectively. 
For example, in the case of $m_{1} = 0$, the diagonalized mass matrix $m_{\n}^{\rm diag}$ has the following $U(1)_{L}$ lepton number symmetry for the massless mode \cite{Branco:1988ex}
\begin{align}
(m_{\n}^{\rm diag})' = R_{1} m_{\n}^{\rm diag} R_{1} \equiv \Diag{e^{i \th}}{1}{1} \Diag{0}{m_{2}}{m_{3}} \Diag{e^{i \th}}{1}{1} = m_{\n}^{\rm diag} \, , 
\end{align}
where $R_{1}$ is a phase matrix. Returning to the basis of original $m_{\n}$ by $U$, 
\begin{align}
U R_{1} U^{\dg} m_{\n} U^{*} R_{1} U^{T} &=  m_{\n}  \, .
\end{align}
This $U R_{1} U^{\dg} $ is a unitary matrix such that
\begin{align}
U R_{1} U^{\dg} 
= e^{i \th} \bm u \otimes \bm u^{\dg} + \bm v \otimes  \bm v^{\dg} + \bm w \otimes \bm w^{\dg} 
= 1 - (1- e^{i\th}) \bm u \otimes \bm u^{\dg} \, . 
\end{align}
Similarly, for $m_{3} = 0$, there is a $U(1)_{L}$ symmetry by $U R_{3} U^{\dg} = 1 - (1- e^{i\th}) \bm w \otimes \bm w^{\dg}$.
Since this chiral symmetry is not necessarily a symmetry of the entire theory,  it is treated as a kind of remnant symmetry. 

The condition that $m_{\n}$ is invariant under the symmetry can be rewritten as
\begin{align}
(1 - P + e^{i\th} P ) \, m_{\n} \, (1 - P + e^{i\th} P)^{T} & = m_{\n} \, ,
\label{U1}
\end{align}
where $P = \bm u \otimes \bm u^{\dg}$ is a projection to the eigenvector of the massless mode $\bm u = \bm u^{\rm NH}$ or $\bm u^{\rm IH}$. 
In order for this symmetry to hold for any $\theta$, the following three equations must hold independently;
\begin{align}
& (1- P) m_{\n} (1-P)^{T} = m_{\n} \, ,  \\
& e^{i\th} (1 - P  ) m_{\n} P^{T} + e^{i\th} P  m_{\n} (1 - P )^{T}   = 0 \, ,  \label{msym} \\
& e^{2 i\th} P m_{\n}  P^{T} = 0 \, .
\end{align}
Since $m_{\n}$ is a symmetric matrix, the condition $e^{i\th} (1 - P ) m_{\n} P^{T} = 0$ is sufficient for Eq.~(\ref{msym}).
Adding and subtracting these three conditions lead to
\begin{align}
(1- P) m_{\n} = m_{\n} (1-P)^{T} = m_{\n} \, , ~~ P  m_{\n} = m_{\n}  P^{T} = 0 \, . 
\end{align}
In other words,  a massless mode associated with $\bm u$ appears if $m_{\n}$ does not have a projective component of this direction. In particular, since $\bm u$ is not the zero vector $\bm 0$, 
\begin{align}
 m_{\n} P^{T} = m_{\n} \bm u^{*} \otimes \bm u^{T} = 0 ~~ \Getsto ~~ m_{\n} \bm u^{*} = \bm 0 \, . 
 \label{cond}
\end{align}

In the type-I seesaw mechanism, this condition becomes
\begin{align}
P \, m_{D} M_{R}^{-1} m_{D}^{T} =  m_{D} M_{R}^{-1} m_{D}^{T} P^{T} = 0  \, .
\end{align}
The existence of $M_{R}^{-1}$ with $\det M_{R} \neq 0$ implies $P \, m_{D} = 0$ and $m_{D}^{T} \bm u^{*} =\bm 0$.
That is, $m_{D}$ also respects the chiral $U(1)$ symmetry due to the left-handed lepton number\footnote{The anti-symmetry $(1 - P + e^{i\th} P ) \, m_{D} = - m_{D}$ has only the trivial solution $m_{D} = 0$.}, 
\begin{align}
(1 - P + e^{i\th} P ) \, m_{D} & = m_{D} \, ,
\label{U2}
\end{align}
and $m_{D}$ has no projection in the eigenvector of zero mode.
We will show later that the same condition holds for $\det M_{R} = 0$.
 
The necessary condition $(P m_{\n} = 0 \To P m_{D} = 0)$ also holds when $\det M_{R} \neq 0$ is satisfied. 
By assuming $m_{1} = 0$ for simplicity, 
SVDs of two matrices (with rank two) $m_{\n} = U m_{\n}^{\rm diag} U^{T}$ and $m_{D} = U_{D} m_{D}^{\rm diag} V_{D}^{\dg}$ lead to 
\begin{align}
U \Diag{0}{m_{2}}{m_{3}} U^{T} 
= U_{D} \Diag{0}{m_{D2}}{m_{D3}} V_{D}^{\dg} M_{R}^{-1} V_{D}^{*} 
\Diag{0}{m_{D2}}{m_{D3}} U_{D}^{T} \, ,
\end{align}
where $m_{D(2,3)}$ are singular values of $m_{D}$. 
Performing production of matrices between two $m_{D}^{\rm diag}$, we obtain
\begin{align}
 \Diag{0}{m_{2}}{m_{3}} 
= U^{\dg} U_{D} 
\begin{pmatrix}
0 & 0& 0 \\
0 & * & * \\
0 & * & *
\end{pmatrix}
 U_{D}^{T} U^{*} \, ,
\end{align}
where $*$ denotes any matrix element. 
Since this is also a SVD of $m_{\n}$, 
$U^{\dg} U_{D}$ must be a unitary transformation of the 2-3 submatrix;
\begin{align}
U^{\dg} U_{D} = 
\begin{pmatrix}
1 & 0 & 0 \\
0 & * & * \\
0 & * & * \\
\end{pmatrix} . 
\end{align}
Therefore the eigenvectors of zero modes must be identical, 
and $P m_{\n} = 0 \To P m_{D} = 0$ holds.
From this, the existence of this chiral symmetry is a necessary and sufficient condition for $m_{\n}$ and $m_{D}$.

This is due to the nature that the left-handed chiral symmetry gives constraints only for $m_{D}$ and does not depend on $M_{R}$~\cite{Adhikari:2010yt}. 
It is easy to show as follows. 
Since $R_{1} = {\rm diag} (e^{i \th} \, , \, 1 \, , \, 1)$ is the generator of the chiral symmetry in the diagonal basis of $m_{\n}$, 
the matrices satisfying $R_{1} m_{\n, D} = m_{\n , D}$ can be written as
\begin{align}
m_{\n}^{\rm diag} = \Diag{0}{m_{2}}{m_{3}} =  
\begin{pmatrix}
0 &0 & 0 \\ a & b & c \\ d & e & f
\end{pmatrix} \, 
\begin{pmatrix}
 M_{11} & M_{12} & M_{13} \\
 M_{12} & M_{22} & M_{23} \\
 M_{13} & M_{23} & M_{33} \\
\end{pmatrix}^{-1}
\begin{pmatrix}
0 &a  & d \\ 0 & b & e \\ 0 & c & f
\end{pmatrix} \, ,
\label{xx}
\end{align}
with arbitrary parameters $ a \sim f$. 
Although a unitary transformation of $M_{R}$  changes magnitudes of $a \sim f$, 
the chiral symmetry is preserved. 

The minimal seesaw model \cite{Ma:1998zg, King:1998jw, Frampton:2002qc, Xing:2020ald} 
 with $\det M_{R} = 0$  always has this chiral symmetry, 
 because the limit of $M_{33} \to \infty$ removes $M_{i3}$ and $c,f$ from the low energy. 
Specifically, by substituting $M_{3} \to \infty$ or $C_{i} = 0$ in the natural representation~(\ref{formula}), a cross product $\bm A \times \bm B$ represents the eigenvector of the massless mode and generates the  chiral symmetry. 
 
\vspace{12pt}

For $m_{D}$ to satisfy the chiral symmetry~(\ref{U2}), it is sufficient to have this subgroup  of a chiral $Z_{2}$ symmetry with $\th = \pi$~\cite{Yang:2022yqw}. 
If vectors $\bm A$ and $\bm B$ are linearly independent, 
\begin{align}
S m_{D} = +  m_{D} \, , ~~~ S = 1 - 2 \bm u \otimes \bm u^{\dg} \, , ~~ \bm u \equiv {(\bm A \times \bm B)^{*} \over |\bm A \times \bm B|} \, .
\label{Z2}
\end{align}
%
The anti-symmetric condition $S m_{D} = - m_{D} $ is unsuitable because the rank of $m_{D}$ becomes unity. 
Such relations between eigenvectors of $m_{\n}$ and the residual $Z_{2}$ symmetry have been well discussed \cite{Lam:2006wy, Lam:2006wm,Lam:2007qc,Lam:2008rs}. 
Since it is obvious that this chiral $Z_{2}$ symmetry reduces the rank of $m_{D}$ by one, 
this symmetry is a necessary and sufficient condition for the massless lightest neutrino\footnote{One might suspect there exists another solution $S m_{D} S_{R}^{\dg} = m_{D}$ with a symmetry of $M_{R}$, $S_{R}^{T} M_{R} S_{R} = M_{R}$. However, it does not satisfy the chiral symmetry because $S m_{\n} = S m_{D} M_{R}^{-1} m_{D}^{ T} = m_{D} M_{R}^{-1} S_{R}^{*} m_{D}^{T} \neq m_{\n}.$}.

For exapmle, for approximate eigenvectors of massless modes $\bm u^{\rm NH} = {1\over \sqrt 6} (2 \, , -1 \, , -1)^{T}$ and $\bm u^{\rm IH} = {1\over \sqrt 2} (0,1,-1)^{T}$, 
generators of the symmetry $S_{i}$ are
\begin{align}
S_{1} \equiv 1 - 2 \bm u^{\rm NH} \otimes (\bm u^{\rm NH})^{\dg} =   
{1\over 3}
\begin{pmatrix}
-1 & 2 & 2 \\
2 & 2 & -1 \\
2 & -1 & 2 \\
\end{pmatrix} \, , ~~~
S_{3}  \equiv 1 - 2 \bm u^{\rm IH} \otimes (\bm u^{\rm NH})^{\dg} =
\begin{pmatrix}
1 & 0 & 0 \\
0 & 0 & 1 \\
0 & 1 & 0 \\
\end{pmatrix} \, .
\end{align}
Note that $S_{3}$ generates a chiral $\m - \t$ symmetry \cite{Fukuyama:1997ky,Lam:2001fb,Ma:2001mr,Balaji:2001ex}.
For each of these, matrices $m_{D}$ with chiral $Z_{2}$ symmetry are
\begin{align}
m_{D}^{\rm (NH)} =
\begin{pmatrix}
{A_{2} + A_{3} \over 2} &  {B_{2} + B_{3} \over 2 } & {C_{2} + C_{3} \over 2 } \\
 A_{2} & B_{2} & C_{2} \\
 A_{3} & B_{3} & C_{3}
\end{pmatrix} , ~~~ 
m_{D}^{\rm (IH)} = 
\begin{pmatrix}
A_{1} & B_{1} & C_{1} \\
 A_{2} & B_{2} & C_{2} \\
 A_{2} & B_{2} & C_{2}
\end{pmatrix} ,
\label{const}
\end{align}
and the symmetry fix a row of $m_{D}$.

Since $m_{D}^{\rm (NH, IH)}$ has no $\bm u^{\rm (NH, IH)}$ component,  
these textures can be realized by a linear combination of vacuum expectation values of the following flavons; 
\begin{align}
\vev{\phi_{1}} = {1\over \sqrt 6} \Column{2}{-1}{-1} \, , ~~
\vev{\phi_{2}} = {1\over \sqrt 3} \Column{1}{1}{1} \, ,  ~~
\vev{\phi_{3}} = {1\over \sqrt 2} \Column{0}{-1}{1} \, .
\end{align}
From this fact,  a massless neutrino and a simple unification with an up-type Yukawa matrices $Y_{u} \sim Y_{\n}$ seem incompatible because
because $m_{D}$ with exact chiral $Z_{2}$ symmetry have similar sizes of two elements in a certain column.

\section{Chiral perturbation theory for the lightest mass  $m_{1 \, \rm or \, 3}$}

The finite lightest mass $m_{1 \, \rm or \, 3}$ makes Eq.~(\ref{U2}) an approximate chiral lepton number symmetry.
Then, let us survey how the condition  for $m_{D}$ such as~(\ref{const}) is relaxed by a perturbatively light $m_{1 \, \rm or \, 3}$, i.e., ``chiral perturbation theory''\cite{Gasser:1984gg} for $m_{\n}$. 

In the diagonal basis of $M_{R}$, the natural representation~\cite{Barger:2003gt} of the  matrix $m_{\n}$ is
\begin{align}
m_{\n} &= m_{D} M_{R}^{-1} m_{D}^{T} 
=  {1 \over M_{1}} {\bm A} \otimes {\bm A}^{T} + {1 \over  M_{2}}  \bm B \otimes \bm B^{T} + {1 \over M_{3}} \bm C \otimes \bm C^{T} \, ,
\label{formula}
\end{align}
where $M_{i}$ are mass singular values of $M_{R}$. 
In well-considered models, there are two possibilities for the massless mode; 
\begin{enumerate}
\item The limit of $M_{3} \to \infty$. 
\item $\bm C$ is linearly dependent on $\bm A$ and $\bm B$. 
\end{enumerate}
To treat the linear independence of $\bm C$ from $\bm A$ and $\bm B$ as perturbative parameters, $m_{D}$ and $m_{\n}$ are devided as follows; 
%
\begin{align}
m_{D} &= m_{D 0} + \d m_{D} 
\equiv  (\bm A \, , \bm B \, , \bm C_{0}) +  (\bm 0 \, , \bm 0 \, ,  \d \bm C) \, ,  \\
 & ~~~   (\bm A \times \bm B)^{T} \bm C_{0} = \bm A^{\dg} \d \bm C =  \bm B^{\dg} \d \bm C = 0 \, ,   \label{ortho} \\
m_{\n} & 
= (m_{D 0} + \d m_{D} ) M_{R}^{-1} (m_{D 0} + \d m_{D} )^{T}
\equiv  m_{\n 0} + \d m_{\n} + \d^{2} m_{\n} \nn \\
& \equiv m_{D 0} M_{R}^{-1} m_{D 0}^{T} + {1\over M_{3}} (\bm C_{0} \otimes \d \bm  C^{T} + \d \bm C \otimes \bm C_{0}^{T}) + {1\over M_{3}}  \d \bm C \otimes \d \bm C^{T}  \, . \label{mndef}
\end{align}
In Eq.~(\ref{ortho}), note that $\bm A \, , \bm B \,$ and $(\bm A \times \bm B)^{*}$ form a basis under the Hermitian inner product. 
Although other parameterizations of the breaking are possible, 
the expression is significant because the inverse of the heaviest  mass $M_{3}^{-1}$ is associated with the parameters and we can use the fact that $\bm C$ is hierarchical ($|C_{3}| \gg |C_{1,2}|$)
 in many models.

The unitary matrix $U$ that diagonalizes $m_{\n}$ is divided as 
\begin{align}
U = U_{0} + \d U \equiv (\bm u_{0} \, , \, \bm v_{0} \, , \, \bm w_{0}) + (\d \bm u \, , \, \d \bm v \, , \, \d \bm w) \, , 
\end{align}
where $\bm u_{0} \propto (\bm A \times \bm B)^{*} $ and 
$\bm v_{0}, \bm w_{0}$ are linear combinations of $\bm A$ and $\bm B$. 
The second-order perturbation $\d^{2} U$ is not considered 
because it does not contribute to the lowest-order calculation. 

By considering NH for simplicity, $m_{\n}$ in the basis of diagonalizing $m_{\n 0}$ is
\begin{align}
U_{0}^{\dg} m_{\n} U_{0}^{*} 
= \Diag{0}{m_{2}}{m_{3}} + {1\over M_{3}}
\begin{pmatrix}
C_{u}^{2} & C_{u} C_{v} & C_{u} C_{w}  \\
C_{u} C_{v} & 0 & 0 \\
C_{u} C_{w} &0 & 0
\end{pmatrix} , 
\label{pertbd}
\end{align}
where
$C_{u} \equiv \bm u_{0}^{\dg} \bm C = \bm u_{0}^{\dg} \d \bm C$ and  $C_{v} \equiv  \bm v_{0}^{\dg} \bm C = \bm v_{0}^{\dg} \bm C_{0}$ by Eq.~(\ref{ortho}).  
A similar notation is used for $\bm w_{0}$. 
The parameter $C_{u}$ represents a magnitude of chiral perturbation
because $C_{u} \propto (\bm A \times \bm B)^{T} \bm C = \det m_{D}$.
Thus, the change of eigenvector $\d \bm u$ is a first-order perturbation, and
the lightest mass value $m_{1}$ arises from a second-order perturbation.
The detailed calculation is found in the appendix.

What needs to be investigated is how the constraints such as~(\ref{const}) are shifted by 
the chiral symmetry breaking. 
For example, the deviation $\d \bm u$ for the massless eigenvector $\bm u_{0}^{*} \propto \bm A \times \bm B$ is evaluated as follows. 
Performing a further diagonalization for Eq.~(\ref{pertbd}), 
we obtain the deviation from the non-perturbed unitary matrix $U_{0}$ as 
\begin{align}
1 + U_{0} \d U^{\dg} \simeq 
\begin{pmatrix}
1 &-  { C_{u} C_{v} \over m_{2} M_{3}} & - { C_{u} C_{w} \over m_{3} M_{3}}  \\ 
{ C_{u} C_{v} \over m_{2} M_{3}}  & 1 & 0 \\
 { C_{u} C_{w} \over m_{3} M_{3}}  & 0 & 1
\end{pmatrix} = 1 +  
\Row{\bm u_{0}}{\bm v_{0}}{\bm w_{0}}
\Column{~~~~ \d \bm u^{\dg} ~~~~ }{\d \bm v^{\dg}}{\d \bm w^{\dg}}, 
\label{29}
\end{align}
\begin{align}
\bm v_{0}^{\dg} \d \bm u =  - {C_{u}^{*} C_{v}^{*} \over m_{2}M_{3} } \, , ~~~
\bm w_{0}^{\dg} \d \bm u =  - {C_{u}^{*} C_{w}^{*} \over m_{3} M_{3} } \, . 
\label{30}
\end{align}
This evaluation is also obtained by substituting $j=1$ and 
the unperturbed lightest mass $m_{01} = 0$ into the expression~(\ref{dU}) 
 of the SVD.
 %
%

We estimate approximate magnitudes of these parameters. 
Since the elements $(m_{\n 0})_{22}$ and $(m_{\n 0})_{33}$ 
have terms $C_{v}^{2}/M_{3}$ and $C_{w}^{2}/M_{3}$, 
$C_{v }$ and $C_{w}$ have upper bound about $\sqrt{m_{2} M_{3}}$ and $ \sqrt{m_{3} M_{3}}$
if there is no fine-tuning between the sums in $m_{\n 0}$. 
%
Since we will see later $C_{u} \sim \sqrt{m_{1} M_{3}}$, 
the perturbations in Eq.~(\ref{29}) are suppressed by at least $\sqrt{m_{1} /m_{2}}$ and $\sqrt{m_{1} /m_{3}}$ respectively.

The constraint for $m_{D}$~(\ref{const}) is changed 
by the perturbed eigenvector of the lightest mode.
For simplicity, let $U_{0}$ be the tri-bi-maximal mixing \cite{Harrison:2002er} and ignore the contribution of $\bm w_{0}^{\dg} \d \bm u$ in Eq.~(\ref{30}) by $m_{3} \sim 6 \, m_{2}$.
The eigenvector $\bm u$ actually observed is
\begin{align}
\bm u = \bm u_{0} + \d \bm u \simeq {1 \over \sqrt 6} \Column{2}{-1}{-1} - {C_{u}^{*} C_{v}^{*} \over m_{2} M_{3} } 
{1\over \sqrt 3} \Column{1}{1}{1} \, .
\end{align}
Conversely this means that $\bm A$ and $\bm B$ have $\bm u$ components that are suppressed by at least $O (\sqrt{m_{1}/m_{2}})$; 
\begin{align}
\bm u^{\dg} \bm A \simeq - {C_{u} C_{v} \over m_{2} M_{3} } 
\bm v^{\dg} \bm A \, , ~~~
\bm u^{\dg} \bm B \simeq - {C_{u} C_{v} \over m_{2} M_{3}} 
\bm v^{\dg} \bm B \, . 
\label{correction}
\end{align}

Next,  the lightest mass $m_{1}$ is found to be Eq.~(\ref{m1-2}) 
from the detailed calculation of the second-order perturbation; 
\begin{align}
m_{1} = \abs {{ C_{u}^{2} \over M_{3}} \lsp 1  - { C_{v}^{2} \over m_{2} } - { C_{w}^{2} \over m_{3} } \rsp } \, . 
\label{m1}
\end{align}
Intuitively, Eq.~(\ref{m1}) agrees with an approximate evaluation of the perturbative contribution like the seesaw mechanism in Eq~(\ref{pertbd}).
A relation $C_{u} \sim \sqrt{m_{1} M_{3}}$ holds 
because $C_{v}$ and $C_{w}$ has an upper bound around
$\sqrt{m_{2} M_{3}}$ and $\sqrt{m_{3} M_{3}}$. 
In particular, $m_{1}$ is proportional to $\det m_{D}^{2} / M_{3}$ because of $C_{u} = \bm u_{0}^{\dg} \bm C \propto \det m_{D}$. 

Moreover, $|C_{1, 2}| \ll |C_{3}|$ is expected in many unified theories.
If mixings in $U_{0}$ are as large as the MNS matrix in this hierarchical basis, 
a rough expression $\bm C \sim (0 \, , 0 \, , 1)^{T}$ leads to $|C_{u}| \sim |C_{v}| \sim |C_{w}| \sim \sqrt{m_{1} M_{3}}$. 
Therefore, $m_{1}$ in Eq.~(\ref{m1}) is approximately equal to the first term $|C_{u}^{2} / M_{3}|$
and the order of $\d \bm u$ in Eq.~(\ref{30}) is reduced to $m_{1}/ m_{2,3}$ instead of $\sqrt {m_{1}/ m_{2,3}}$.
Conditions that perturbations are sufficiently small in such a model is
\begin{align}
m_{1}^{\rm NH} \lesssim  1\, \meV \, , ~~~
m_{3}^{\rm IH} \lesssim  5 \, \meV \, . 
\end{align}
In this case, corrections~(\ref{correction}) to the constraints for the Dirac mass matrix~(\ref{const}) is about 10\% or less for the lighter (the first and second) generations.

In such models with large $C_{3} = (m_{D})_{33}$, although the lepton number symmetry~(\ref{U1}) and chiral symmetry~(\ref{U2}) are largely broken, 
there still remains a partial chiral ($Z_{2}$) symmetry~\cite{Yang:2022lyu} such that
\begin{align}
S m_{D} P_{3} = +  m_{D} P_{3} \, ,  ~~~ P_{3} \equiv {\rm diag} ( 1 \, , \, 1 \, , \, 0) \, .
\end{align}
Equivalently, for $\bm u^{\rm NH} = {1\over \sqrt 6} (2 \, , -1 \, , -1)^{T}$ and $\bm u^{\rm IH} = {1\over \sqrt 2} (0,1,-1)^{T}$, the constraints are 
\begin{align}
m_{D}^{\rm (NH)} \simeq 
\begin{pmatrix}
{A_{2} + A_{3} \over 2} &  {B_{2} + B_{3} \over 2 } & C_{1} \\
 A_{2} & B_{2} & C_{2} \\
 A_{3} & B_{3} & C_{3}
\end{pmatrix} , ~~~ 
m_{D}^{\rm (IH)} \simeq 
\begin{pmatrix}
A_{1} & B_{1} & C_{1} \\
 A_{2} & B_{2} & C_{2} \\
 A_{2} & B_{2} & C_{3}
\end{pmatrix} .
\end{align}
Since $\bm A$ and $\bm B$ must be approximately orthogonal to the eigenvectors of the massless mode, 
two complex parameters in the lighter generations are constrained.

Finally, such (partial) chiral symmetries are hardly renormalized and retained good symmetries \cite{tHooft:1979bh}.
This property holds as long as a renormalization group equation of a Yukawa matrix $Y_{\n}$ is proportional to the Yukawa itself $dY_{\n} /dt \propto Y_{\n} \simeq S Y_{\n}$. 

\section{Summary}

In this paper, we perform chiral perturbative analysis of an approximate lepton number symmetry associated with a sufficiently light neutrino in the type-I seesaw mechanism. 
For the Dirac mass matrix $m_{D} = (\bm A \, , \bm B \, , \bm C)$, linearly independent components of $\bm C$ from $\bm A$ and $\bm B$ are treated as  symmetry-breaking parameters. 
A deviation in the eigenvector of the massless mode $\d \bm u$ 
occurs in the first-order perturbation and 
the lightest mass $m_{1 \, \rm or \, 3} \propto \det m_{D}^{2} / M_{3}$ emerges in the second-order. 
By solving the perturbation theory,  
we obtained specific expressions of $\d \bm u$ and $m_{1 \, \rm or \, 3}$. 

If $m_{D}$ satisfies the hierarchy $|(m_{D})_{13}| , |(m_{D})_{23}| \ll |(m_{D})_{33}|$ and the diagonalization of the mass matrix of light neutrinos $m_{\n}$ has large mixing in this hierarchical basis, the order of $\d \bm u$ is reduced to $m_{1 \, \rm or \, 3} / m_{2}$.
Thus the perturbative description works well for $m_{1}^{\rm NH} \lesssim 1\, \meV \, , ~ m_{3}^{\rm IH} \lesssim 5 \, \meV$.
As a result, in the type-I seesaw mechanism with a hierarchical Dirac mass matrix, 
flavor structures of the lighter generation are approximately orthogonal to the eigenvector of the lightest neutrino, and two complex parameters for $m_{D}$ are constrained from the eigenvector.
 
\section*{Acknowledgment}
This study is financially supported 
by JSPS Grants-in-Aid for Scientific Research
No.~JP18H01210 and MEXT KAKENHI Grant No.~JP18H05543.

\appendix 

\section{Detailed calculation of chiral perturbation theory}

This section shows a detailed calculation 
of the chiral perturbation for the SVD of $m_{\n}$.
If the three neutrino masses $m_{i}$ are finite, $\bm A, \bm B$ and $\bm C$ are linearly independent. To treat this linear independence of $m_{D}$ as perturbations, $m_{D}$ and $m_{\n}$~(\ref{formula}) can be divided as follows; 
\begin{align}
m_{D} &= m_{D 0} + \d m_{D} 
\equiv  (\bm A \, , \bm B \, , \bm C_{0}) +  (\bm 0 \, , \bm 0 \, , \d \bm C) \, ,  \\
 & ~~~   (\bm A \times \bm B)^{T} \bm C_{0} = \bm A^{\dg} \d \bm C =  \bm B^{\dg} \d \bm C = 0 \, ,   \label{ortho2} \\
m_{\n} & 
= (m_{D 0} + \d m_{D} ) M_{R}^{-1} (m_{D 0} + \d m_{D} )^{T}
\equiv  m_{\n 0} + \d m_{\n} + \d^{2} m_{\n} \nn \\
& \equiv m_{D 0} M_{R}^{-1} m_{D 0}^{T} + {1\over M_{3}} (\bm C_{0} \otimes \d \bm  C^{T} + \d \bm C \otimes \bm C_{0}^{T}) + {1\over M_{3}}  \d \bm C \otimes \d \bm C^{T}  \, . \label{mndef2}
\end{align}
%
%
It is reasonable to treat $\d \bm C$  as symmetry-breaking parameters 
because $\det m_{D} = (\bm A \times \bm B)^{T} \d \bm C$ holds. 
%

The unitary matrix $U$ that diagonalizes $m_{\n}$ is divided as 
\begin{align}
U = U_{0} + \d U \equiv (\bm u_{0} \, , \, \bm v_{0} \, , \, \bm w_{0}) + (\d \bm u \, , \, \d \bm v \, , \, \d \bm w) \, , 
\end{align}
where $\bm u_{0} \propto (\bm A \times \bm B)^{*} $ and 
$\bm v_{0}, \bm w_{0}$ are linear combinations of $\bm A$ and $\bm B$. 
The second-order perturbation $\d^{2} U$ is not considered 
because it does not contribute to the lowest-order calculation\footnote{Even if we consider $\d^2 U$, terms like $U_{0} m_{\n 0} \d^{2} U$ do not contribute to $(\d^{2} m_{\n}^{\rm diag})_{11}$.}.

The SVDs are given by
\begin{align}
U^{\dg}_{0} m_{\n 0} U_{0}^{*} = m_{\n 0}^{\rm diag} \, ,   ~~~ 
U^{\dg} m_{\n} U^{*} &= m_{\n}^{\rm diag} = m_{\n 0}^{\rm diag} + \d m_{\n}^{\rm diag} 
+ \d^{2} m_{\n}^{\rm diag} \, .
\label{diag}
\end{align}
The diagonalization of the first-order perturbation to $m_{\n} m_{\n}^{\dg}$ is
\begin{align}
(U_{0} + \d U)^{\dg} (m_{\n 0} + \d m_{\n} )  (m_{\n 0}^{\dg} + \d m_{\n}^{\dg} )(U_{0} + \d U) & = (m_{\n 0}^{\rm diag})^{2} + 2 m_{\n 0}^{\rm diag} \d m_{\n}^{\rm diag} + (\d m_{\n}^{\rm diag})^{2} \, . 
\end{align}
By using the diagonalization~(\ref{diag}) and the orthogonality relation $\d U^{\dg}\, U_{0} + U^{\dg}_{0} \d U = 0$, 
\begin{align}
 & U_{0}^{\dg} m_{\n 0} m_{\n 0 }^{\dg} \d U + \d U ^{\dg }m_{\n 0} m_{\n 0 }^{\dg} U_{0} + U_{0}^{\dg} m_{\n 0} \d m_{\n}^{\dg} U_{0} + U_{0}^{\dg} \d m_{\n} m_{\n 0}^{\dg}  U_{0} = 2 m_{\n 0}^{\rm diag} \d m_{\n}^{\rm diag} \, ,  \\
= & - (m_{\n 0}^{\rm diag})^{2} \d U ^{\dg} U_{0} + \d U ^{\dg } U_{0} (m_{\n 0}^{\rm diag})^{2} + m_{\n 0}^{\rm diag} U_{0}^{T} \d m_{\n}^{\dg} U_{0} + U_{0}^{\dg} \d m_{\n} U_{0}^{*} m_{\n 0}^{\rm diag} \, . \label{26}
\end{align}
For the diagonal element of Eq.~(\ref{26}), 
\begin{align}
 (U_{0}^{T} \d m_{\n}^{\dg} U_{0} + U_{0}^{\dg} \d m_{\n} U_{0}^{*})_{ii} =  2  (\d m_{\n}^{\rm diag} )_{ii} \, .
\label{deltami}
\end{align}
In particular, $(\d m_{\n}^{\rm diag} )_{11} = 0$ holds, and the lightest mass occurs in the second-order perturbation. For the off-diagonal elements of Eq.~(\ref{26}) with $i \neq j$, 
\begin{align}
 - m_{0 i }^{2} (\d U ^{\dg} U_{0})_{ij} + (\d U ^{\dg } U_{0})_{ij} m_{0 j}^{2} + m_{0 i}  (U_{0}^{T} \d m_{\n}^{\dg} U_{0})_{ij} + (U_{0}^{\dg} \d m_{\n} U_{0}^{*})_{ij} m_{0 j} = 0_{ij} \, , 
\end{align}
where $m_{0 i} = (m_{\n 0}^{\rm diag})_{i}$.
Thus, we obtain
\begin{align}
 ( U_{0}^{\dg} \d U)_{ij} & = 
- { m_{0 i}  (U_{0}^{T} \d m_{\n}^{\dg} U_{0})_{ij} +  (U_{0}^{\dg} \d m_{\n} U_{0}^{*})_{ij} m_{0 j} \over m_{0 i}^{2} - m_{0 j }^{2} }  \, . 
\label{dU}
\end{align}
From $U_{0}^{\dg} ( U_{0} + \d U) = 1 + U_{0}^{\dg} \d U$, 
it represents a perturbative transformation in the diagonalized basis of $m_{\n 0}^{\rm diag}$.  \\

Next, to estimate the lightest mass $m_{1 \, \rm or \, 3}$, 
the perturbation theory proceeds to the second order. 
For simplicity, we directly consider a diagonalization of $m_{\n}$ instead of $m_{\n} m_{\n}^{\dg}$; 
\begin{align}
(U_{0} + \d U)^{\dg} (m_{\n 0} + \d m_{\n} + \d^{2} m_{\n} ) (U_{0} + \d U)^{*} & = m_{\n 0}^{\rm diag} + \d m_{\n}^{\rm diag} + \d^{2} m_{\n}^{\rm diag} \, ,  \\
\d U^{\dg} m_{\n 0} U_{0}^{*} + U_{0}^{\dg} \d m_{\n} U_{0}^{*} + 
U_{0}^{\dg} m_{\n 0} \d U^{*}  & =  \d m_{\n}^{\rm diag} \, , \label{dm1} \\
U_{0}^{\dg} \d^{2} m_{\n} U_{0} ^{*} +
U_{0}^{\dg}  \d m_{\n}  \d U^{*} +
 \d U^{\dg} m_{\n 0} \d U^{*} +
 \d U^{\dg}  \d m_{\n} U_{0} ^{*} & = \d^{2} m_{\n}^{\rm diag} \, . \label{dm2}
\end{align}
The sum of Eq.~(\ref{dm1}) and its complex conjugate is equivalent to Eqs.~(\ref{deltami}). 
By substituting Eq.~(\ref{dm1}) into Eq.~(\ref{dm2}) and eliminating $\d m_{\n}$, 
\begin{align}
 & U_{0}^{\dg} \d^{2} m_{\n} U_{0} ^{*} +
(  \d m_{\n}^{\rm diag} - m_{\n 0}^{\rm diag} U_{0}^{T} \d U^{*}) U_{0}^{T} \d U^{*} \nn
\\ & -  \d U^{\dg} U_{0} m_{\n 0}^{\rm diag} U_{0}^{T} \d U^{*} +
 \d U^{\dg} U_{0} (  \d m_{\n}^{\rm diag} -\d U^{\dg} U_{0} m_{\n 0}^{\rm diag} ) = \d^{2} m_{\n}^{\rm diag} \, . 
\end{align}
Substituting $(m_{\n 0}^{\rm diag})_{11} = (\d m_{\n}^{\rm diag})_{11} = 0$ to the 1-1 element, we obtain 
\begin{align}
m_{1} & = 
(U_{0}^{\dg} \d^{2} m_{\n} U_{0} ^{*} -  \d U^{\dg} U_{0} m_{\n 0}^{\rm diag} U_{0}^{T} \d U^{*})_{11} \nn \\  
& =  (U_{0}^{\dg} \d^{2} m_{\n} U_{0} ^{*})_{11} - \sum_{i} {  (U_{0}^{\dg} \d m_{\n} U_{0}^{*})_{1i} \over  m_{0 i } } m_{0 i} {  (U_{0}^{\dg} \d m_{\n} U_{0}^{*})_{1i} \over  m_{0 i } } \, . 
\end{align}
%
Finally, $\d m_{\n}$ and $\d^{2} m_{\n}$ in Eq.~(\ref{mndef2}) yields $( U_{0}^{\dg} \d m_{\n} U_{0}^{*} )_{1i}  = {1\over M_{3}} (\bm u_{0}^{\dg}  \d \bm C ) \cdot  (\bm C_{0}^{T} (\bm u_{0}^{*} \, , \bm v_{0}^{*} \, , \bm w_{0}^{*}))_{i}$ and the lightest mass is found to be
\begin{align}
m_{1} = { C_{u}^{2} \over M_{3}} \lsp 1  - { C_{v}^{2} \over m_{02} } - { C_{w}^{2} \over m_{03} }  \rsp \, . 
\label{m1-2}
\end{align}
The subscript 1 is due to the fact that $\bm u_{0}$ was specified as the first column, and it is also valid for IH by a rearrangement to the third column.
The lightest singular value will be the absolute value of this expression 
because the phase can be eliminated by redefining the basis (Eq.~(\ref{deltami}) is the form of conjugate addition because it is a correction to finite singular values. The absolute value should be taken for $m_{1}$ since this is the leading term).
%


\end{document}